\documentstyle[12pt,
epsf]{article}
\newcommand{\hepph}[1]{{\tt hep-ph/#1 }}
\newcommand{\hepex}[1]{{\tt hep-ex/#1 }}
\newcommand{\prd}[3]{{{\it Phys.~Rev.}~{\bf D#1} (#3) #2}}                  
\newcommand{\plb}[3]{{{\it Phys.~Lett.}~{\bf B#1} (#3) #2}}
\newcommand{\npb}[3]{{{\it Nucl.~Phys.}~{\bf B#1} (#3) #2}}
\newcommand{\prl}[3]{{{\it Phys.~Rev.~Lett.}~{\bf #1} (#3) #2}}

\newcommand{\smallfrac}[2]{\frac{\mbox{\small #1}}{\mbox{\small #2}}}
\newcommand{\leqsim}{\,\raisebox{-0.6ex}{$\buildrel < \over \sim$}\,}
\newcommand{\geqsim}{\,\raisebox{-0.6ex}{$\buildrel > \over \sim$}\,}
\newcommand{\be}{\begin{equation}}
  \newcommand{\ee}{\end{equation}}
\newcommand{\ba}{\begin{eqnarray}}
  \newcommand{\ea}{\end{eqnarray}}
\newcommand{\nn}{\nonumber}
\newcommand{\ie}{\mbox{\em i.e.~}}
\newcommand{\eg}{\mbox{\em e.g.~}}
\def\gev{\,\mbox{GeV~}}
\begin{document}
\baselineskip 18pt
\newcommand{\raltitle}
{ The MSSM at the low $\tan \beta$ fixed point is meta-stable} 
\newcommand{\ralauthor}
{S.A. Abel$^1$, B.C. Allanach$^2$}
\newcommand{\raladdress}
{1. Theory Division, CERN, 1211 Geneva 23 \\
  2. Rutherford Appleton Laboratory, Chilton, Didcot, Oxon. OX11 0QX, \\United
  Kingdom }

\newcommand{\ralabstract}
{We analyse the mass spectrum of the Constrained Minimal 
  Supersymmetric Standard Model at the low $\tan\beta$ fixed point.
  We find that the model 
  only satisfies experimental and dark matter bounds in regions 
  where the vacuum is meta-stable -- \ie where it violates 
  `unbounded from below' (UFB) bounds. Adding a small amount of 
  $R$-parity violation solves these problems but the absolute upper bound
  on the lowest higgs mass   $m_{h^0}<97$ GeV remains. 
We present the predicted sparticle mass spectrum as a function of the gluino
mass $m_g$.
  }

\begin{titlepage}
  \begin{flushright}
    hep-ph/9803476\\
  \end{flushright}
  \vspace{.1in}
  \begin{center}
    {\large{\bf \raltitle}}
    \bigskip \\ \ralauthor \\
    \vspace{\baselineskip} 
    \raladdress \\
    \mbox{} \\ \vspace{.2in}
    {\bf Abstract} \bigskip 
  \end{center} 
  \setcounter{page}{0}
  \ralabstract
\end{titlepage}

\section*{Introduction}
Fixed point behaviour (or rather `quasi-fixed') is a 
striking feature in the Minimal Supersymmetric Standard Model 
(MSSM)~\cite{topfps,others,waoi}. Broadly speaking, it is a focusing 
of the parameters in the infra-red regime which occurs when the 
top quark Yukawa coupling, $h_t$, is large. 
Its existence has been examined in detail
for $h_t$ itself in Refs.\cite{others,waoi} where it was found that
$h_t(m_t)=1.1$ independently of $h_t(M_{GUT})$ 
as long as it is big, say bigger than about 1.2, at the GUT scale.
There are three parameters which always have 
quasi-fixed points (QFPs) regardless of the pattern of supersymmetry breaking, 
and which are strongly attracted towards them (although others formally 
have fixed points as well)~\cite{waoi,as};
\ba
R &\equiv & h_t^2/g_3^2 \nonumber \\
A_t &\equiv & A_{U_{33}} \nonumber \\
3 M^2 & \equiv & m_{U_{33}}^2 + 
m_{Q_{33}}^2 +m_{2}^2.
\ea
When $h_t$ is high at the GUT or Planck scale,
these three parameters are completely determined at the weak scale;
\ba
\label{fps}
R^{QFP} &=& 0.87 \nonumber \\
A_t^{QFP} &=& -1.60 M_{1/2} \nonumber \\
(M^2)^{QFP} &=& 1.83 M_{1/2}^2.
\ea
They govern the running of the MSSM at low $\tan \beta$~\cite{topfps}
and indeed all of the soft supersymmetry breaking parameters may 
easily be solved (to one loop) in terms of them. 
In the appendix, we list the solutions for the running
MSSM mass parameters in terms of the GUT scale values and 
these three parameters. 
Writing the solutions in this form is particularly useful 
for finding combinations of parameters which have QFPs 
with various patterns of supersymmetry breaking including  
non-universal GUT scale conditions 
(\eg in Carena and Wagner of ref.~\cite{others}).
Many of these combinations are flavour off-diagonal 
and first and second generation which leads 
to a natural reduction in FCNCs at low $\tan\beta$~\cite{waoi}. 

At large $h_t$, therefore, quasi-fixed 
behaviour pervades the entire renormalisation group running of the 
MSSM, offering the possibility of a considerably 
reduced parameter space. Moreover 
it was also observed that the recent precise
determination of $m_t=175 \pm 5$\gev~\cite{topmass} means that the top Yukawa
coupling must be large at the GUT scale ($M_{GUT}\sim 10^{16}$ GeV) in the low
$\tan \beta < 30$ regime~\cite{waoi,espin}. 
If $m_t$ lies within (or above) these quoted
1$\sigma$ errors, then quasi-fixed behaviour is indeed going 
to be a dominant feature. It has also been shown numerically~\cite{btau}, 
that bottom-tau Yukawa unification in SUSY GUTs
forces the solutions to be near the QFP. 
(We will explain why with a simple analytic argument below.)
We should stress that quasi-fixed behaviour is a one loop 
effect and in principle it could be destroyed by two loop 
and higher corrections.  
However the quasi-fixed 
behaviour persists to two loop order (including Yukawa 
corrections other than that of $h_t$).

In this letter, we use these predictions to simplify the 
analysis of the mass spectrum for the specific case of the `Constrained' 
MSSM (CMSSM). This is a minimal version of the MSSM with 
the usual $R$-parity invariant MSSM superpotential, 
\be
W_{MSSM}=h_U Q H_2 U^c + h_D Q H_1 D^c + h_E L H_1 E^c+\mu H_1 H_2,
\ee
and a soft SUSY breaking sector which depends
on only four high scale parameters; $A$ (the degenerate trilinear coupling),
$m_0$ (the degenerate scalar mass), $M_{1/2}$ (the degenerate gaugino
mass) and $\tan\beta$ (the ratio of higgs VEVs).
The degeneracy is motivated in part by minimal 
supergravity but we shall, as is usual, impose it at $M_{GUT}$.
The fact that we are working close to the QFP
means that the ratio of higgs VEVs, $\tan \beta$, is fixed by the 
QFP value $h_t(m_t)=1.1$ and the relation
\be
\sin \beta = \frac{m_t(m_t)}{v h_t(m_t)} \label{tanbpred}
\ee
where $m_t(m_t)\approx 160 \pm 5 \gev$ is the $\overline{DR}$ running top
quark mass extracted from
experiment and $v= 174.1 \gev$ is the higgs vacuum expectation value parameter
extracted from $M_Z$. Note that this value of $m_t(m_t)$ is lower than in some
of the literature~\cite{waoi, others} because of the effect of gluino and stop
corrections~\cite{espin}. 
Since the top quark trilinear coupling, $A_t$, also has a fixed point 
and is the trilinear coupling which predominates in the 
mass matrices at low $\tan\beta $, the supersymmetric mass spectrum 
depends only upon $m_0$ and $M_{1/2} $ at the QFP~\cite{others,waoi,espin}. 

We test the spectrum against experimental bounds and
in particular the bound on the lightest higgs.
In addition we consider 
whether there is a charge and/or colour breaking minimum which 
can compete with the physical vacuum~\cite{ccb,komatsu,casas,as}. 
The most restrictive 
constraints come from so called `unbounded from below' (UFB)
directions~\cite{komatsu,casas,as} in which a minimum 
can be generated radiatively essentially because, at 
some point during the running, the mass-squared term 
for $H_2$ must become negative in order to drive 
radiative electroweak symmetry breaking.
The bound which is usually imposed comes from requiring that 
the physical minimum should not be meta-stable. A 
more relevant (and sufficient) condition is to require that there
be no {\em local}\/ minima other than the physical one \cite{as}.
As shown in Ref.\cite{as}, the two conditions are in any case 
numerically very close so we shall use the `traditional' meta-stability
bound. The UFB bounds depend only on $m_0 $ and $M_{1/2}$ 
at the QFP and are expected to be~\cite{casas,as} 
\be
m_0 \geqsim M_{1/2}.
\ee
We then compare the remaining parameter space with that
allowed by dark matter constraints at the QFP~\cite{dn,bargerkao}
and find that the only allowed regions are meta-stable. 
We stress that the MSSM at the low $\tan\beta $ QFP is not yet 
ruled out by Ref.\cite{efos}. There the bounds on $\tan\beta $ 
were $1.4$ and $1.7$ for $\mu<0$ and $\mu>0$ respectively (note 
that we are using the Ref.\cite{bargerkao} definition of the sign 
of $\mu $ which is opposite to that of Ref.\cite{efos}). 
However at the QFP $\tan\beta \approx 1.4\rightarrow 1.5$
for $\mu < 0$, and is largest in the region where $m_0 \leqsim M_{1/2}$,
\ie precisely where the UFB bounds are relevant.
The UFB bounds (which were not included in Ref.\cite{efos})
are therefore an additional and restrictive constraint at the QFP.
(As noted in Ref.\cite{as} they drop quite quickly away 
from the QFP although they are still significant.) 

We finish by discussing how this fact should be interpreted and
also by pointing out that two of these problems (\ie meta-stability 
and the dark matter constraints) can be removed by adding 
$R$-parity violating terms just below experimental bounds~\cite{as}
(albeit at the expense of losing the neutralino as a dark matter 
candidate). 

Before tackling the spectrum, we first expand on the reason why 
Yukawa unification leads to fixed point behaviour. For example,
many SUSY GUTs~\cite{btau} predict the existence of the unification of the
bottom and tau Yukawa couplings at the GUT scale, $\lambda_b
(M_{GUT})=\lambda_\tau (M_{GUT})$; why does this constraint
favour the QFP\@? The RGE for $R_{b/\tau} \equiv 
\lambda_b /\lambda_\tau$ to one loop order is
\be
\frac{d R_{b/\tau}}{d \ln r} = \frac{R_{b/\tau}}{6} \left[ R - 16/3 +
  \frac{4}{3} \frac{\alpha_1}{\alpha_3} \right]
\ee
where, for convenience, we have expressed the running in terms of 
\be
r{\scriptstyle (Q)} \equiv \frac{\alpha_3{\scriptstyle (M_{GUT}})}
{\alpha_3{\scriptstyle (Q)}}=
1-\mbox{\footnotesize 6}\>
{\alpha_3{\scriptstyle (M_{GUT}) }}\log {\scriptstyle (
  \frac{Q}{M_{GUT}})}.  
\ee
The solution is given by 
\be
\frac{R}{R_0} = \left( \frac{R_{b/\tau} (m_t)} {R_{b/\tau} (M_{GUT})}
\right)^{12} r^\frac{89}{9}
\left(\frac{\alpha_2(M_{GUT})}{\alpha_2(m_t)}\right)^3
\left(\frac{\alpha_1(M_{GUT})}{\alpha_1(m_t)}\right)^{\frac{133}{99}},
\label{btconst}
\ee
where $R_0\equiv R {\scriptstyle (M_{GUT}) }$.
As is customary, we define a distance $\rho$ to the QFP,
\be
\rho \equiv 1 - \frac{R}{R^{QFP}} 
= \frac{R}{R_0 \Pi r}, \label{xdist}
\ee
where the last relation can be found, for example, in Ref.\cite{as}, and 
where
\be
\Pi = r^{-16/9}
\left(\frac{\alpha_2(m_t)}{\alpha_2(M_{GUT})}\right)^{-3}
\left(\frac{\alpha_1(m_t)}{\alpha_1(M_{GUT})}\right)^{-13/99}. \label{PI}
\ee
$\rho\ll 1$ near the QFP\footnote{Constraining 
$h_t(M_{GUT})<5$ yields the `perturbativity' condition $R/R_0 >
\frac{1}{56}$ or $\rho > 4\times 10^{-3}$.}. 

Yukawa unification (\ie $R_{b/\tau} (M_{GUT})=1$)
then yields a value for $\rho$ via 
\be
\label{xdist2}
\rho = \left(R_{b/\tau} (m_t)
\right)^{12} r^\frac{32}{3}
\left(\frac{\alpha_1(M_{GUT})}{\alpha_1(m_t)}\right)^{\frac{40}{33}}.
\ee
$R_{b/\tau}(m_t)$ is a number which may be determined from experiment; 
evaluation to three-loop order in QCD and one loop order in
QED yields $R_{b/\tau} =1.48-1.67$ for $\alpha_s(M_Z)$ in the range 0.115-0.121
and $m_b(m_b)$ in the range 4.1-4.4 GeV.
We calculate $\alpha_1(m_t)^{-1}=58.62$, $\alpha_2(m_t)^{-1}=30.022$ from
$\sin^2 \theta_w^{\overline{MS}}=0.2315$, $\alpha(M_Z)^{-1}=127.9$ and
renormalising from $M_Z$ to $m_t$ to one loop accuracy in the 
Standard Model~\cite{benherb}.
Substituting these figures into
Eq.\ref{xdist2} we find
$\rho=7.7 \times 10^{-3} - 2.6 \times 10^{-2}$.
If threshold effects imply~\cite{btau} that $R_{b/\tau}(M_{GUT})=0.9$ or 1.1,
then $\rho$ is 3 times smaller or larger respectively.
In other words bottom-tau Yukawa unification at low $\tan
\beta$ can only be consistent with experiment if the solutions are 
very near to their QFPs. A similar situation holds at high $\tan\beta $.

\section*{The sparticle spectrum and constraints}

\begin{figure}
  \begin{center}
    \leavevmode   
    \hbox{\epsfxsize=15cm
      \epsfysize=15cm
      \epsffile{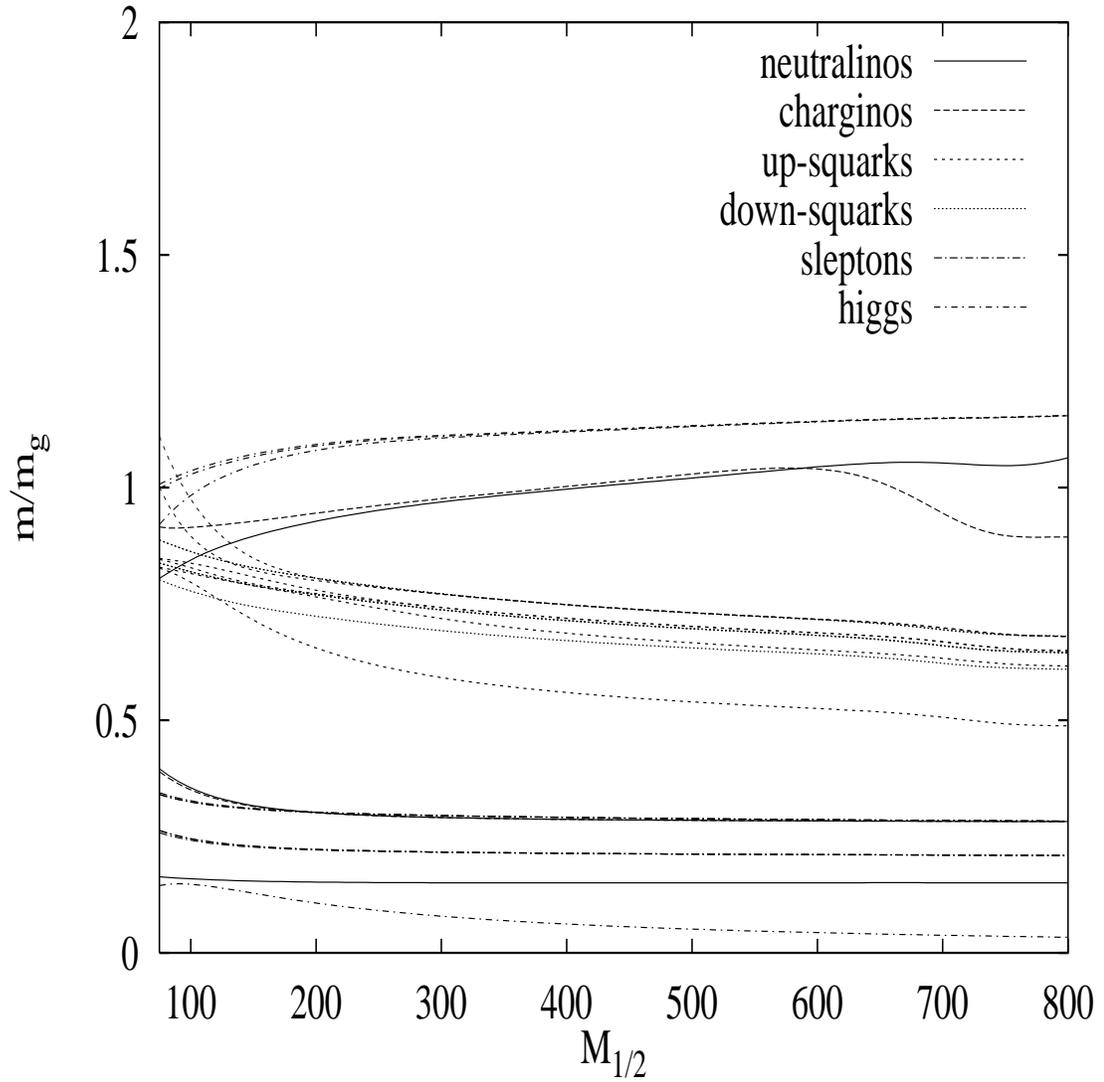}}
  \end{center}
  \caption{The sparticle mass spectrum in the quasi-fixed CMSSM (normalised by
    the gluino mass) vs.\
    $M_{1/2}/$GeV.
    We have chosen the line $m_0= 0.5 M_{1/2}$ of Fig.\protect\ref{fail} and
    $\mu>0$. Note that $m_g\approx 2.7 M_{1/2}$.}
  \label{posspec}
\end{figure}
\begin{figure}
  \begin{center}
    \leavevmode   
    \hbox{\epsfxsize=15cm
      \epsfysize=15cm
      \epsffile{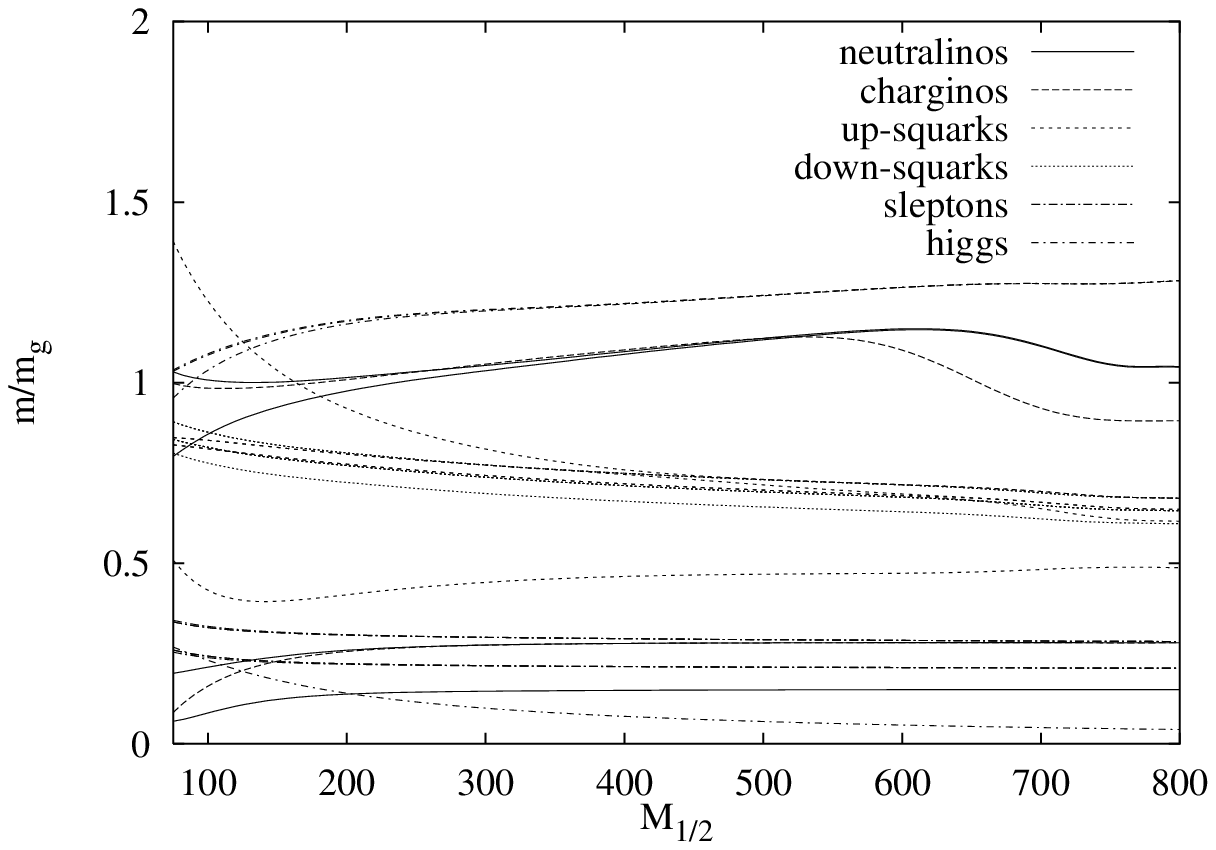}}
  \end{center}
  \caption{As in Fig.\ref{posspec} for $\mu<0$.}
  \label{negspec}
\end{figure}
We now turn to the two loop numerical evaluation of the spectrum. 
In minimal supergravity, the sparticle spectrum depends (generically)
upon the six parameters $\tan \beta$, $A$, $m_0$, $M_{1/2}$, $\mu$, $B$.
The empirically derived value of $m_t$ and the QFP prediction sets 
the first parameter by
Eq.\ref{tanbpred}, and the QFP prediction of $A_t$ eliminates 
the spectrum's dependence upon the second. The parameters were 
run very close to the quasi-fixed point 
(taking $R_0=10$ which corresponds to $\rho=1.8 \times 10^{-2}$) and the full one loop 
potential minimised to determine the higgs couplings, $\mu $ and $B$,
by imposing correct electroweak symmetry breaking. 
(Note that {\em at}\/ the QFP the parameter $\mu $ has a fixed point 
prediction of zero which would be incompatible with electroweak symmetry
breaking.) 
The sign of $\mu$ is retained as an
additional discrete parameter (see Refs.\cite{barger,acw} for details). 
The derivation of $\tan \beta$ ($\sim 1.5$) from $m_t$ was
made using the prescription given in Ref.\cite{Bag}.
Analytic expressions for the light higgs masses may be found 
in Refs.\cite{wagner,haber}; we used those of Ref.\cite{haber} and were 
able to reproduce the figures of Ref.\cite{espin} to within 
$\pm 2\gev $ (although our lightest higgs mass derived using 
a full numerical running mostly fell about 1-2 \gev below that 
in Ref.\cite{espin}).

Given the form of the analytic solutions to the renormalisation group 
equations, the mass spectrum is expected to become proportional to 
$M_{1/2}$ along the line $m_0/M_{1/2}=a$, where $a$ is constant. 
We present the spectrum along the line $m_0=0.5 M_{1/2}$
in Fig.~\ref{posspec} for positive $\mu$ and in Fig.~\ref{negspec}
for negative $\mu$.
It should be noted that the spectrum is generically virtually proportional 
to $M_{1/2}$ along a given line of constant $m_0/M_{1/2}$; 
basically $M_{1/2}$ (or equivalently $m_g$) simply sets the superpartner scale.
The spectrum is found to be almost entirely independent 
of $A$ as expected since the only trilinear coupling entering the
spectrum, $A_t$, has a fixed point given by Eq.\ref{fps}.

The squark/slepton spectrum has a non-trivial dependence at low 
$M_{1/2}$ because $M_Z$ appears in the mass matrices and is 
comparable to $M_{1/2}$
in this region. The heavy neutralinos and charginos  
are dominated by $\mu $ at low values of $M_{1/2}$ until $M_{1/2}$ 
becomes large enough at which point their masses are proportional.
The lightest neutralino and chargino masses are almost
proportional to $M_{1/2}$. In particular we find that the mass of the 
lightest supersymmetric partner lies in the range,
\be
0.15 \leqsim m_{LSP}/m_g \leqsim 0.18
\ee
and agrees well with the empirical analytic approximation 
\ba
m_{\chi_1^0}&\approx 0.448 M_{1/2}+12 \sin 2\beta -10 \mbox{\hspace{1cm} : $\mu > 0 $}\nonumber\\
m_{\chi_1^0}&\approx 0.452 M_{1/2}+5 \sin 2\beta -13 \mbox{\hspace{1cm} : $\mu < 0 $}
\ea 
reported in Ref.\cite{bargerkao}.

\begin{figure}
  \begin{center}
    \leavevmode   
    \hbox{\epsfxsize=15cm
      \epsfysize=15cm
      \epsffile{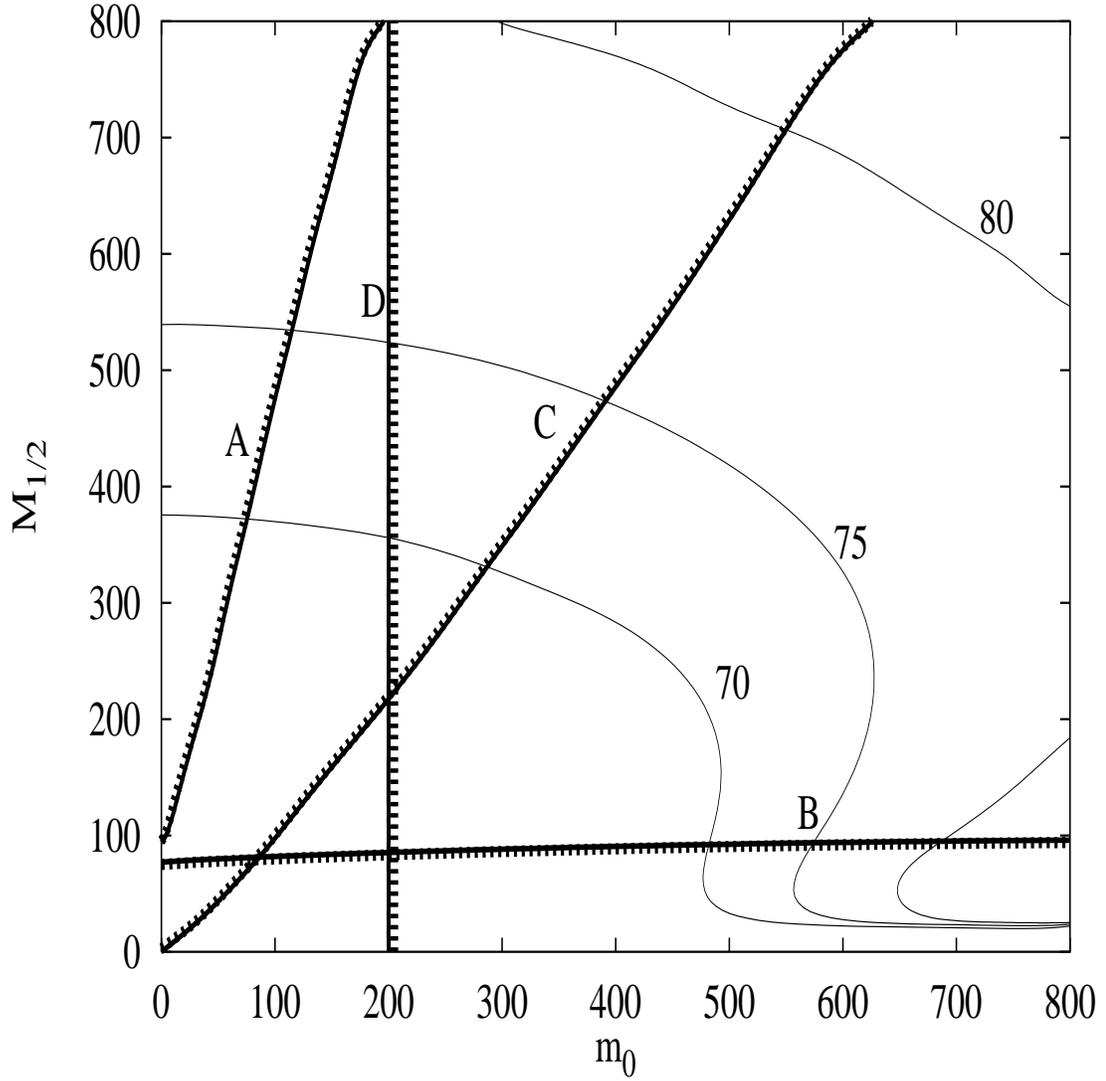}}
  \end{center}
  \caption{Constraints upon the quasi-fixed CMSSM: bounds in the $m_0$
    and $M_{1/2}$ plane for $\mu>0$. 
    $m_0$ and $M_{1/2}$ are measured in GeV. The
    labels correspond to the following requirements: $A$-neutralino is the
    lightest supersymmetric particle (LSP); $B$-chargino mass
    bounds satisfied; $C$-CCB and UFB
    bounds
    satisfied; $D$ non over closure dark matter bound. The lines marked 
    70, 75, 80 GeV give contours of lightest higgs 
    mass $m_{h^0}$.}
  \label{fail} 
\end{figure}
\begin{figure}
  \begin{center}
    \leavevmode   
    \hbox{\epsfxsize=15cm
      \epsfysize=15cm
      \epsffile{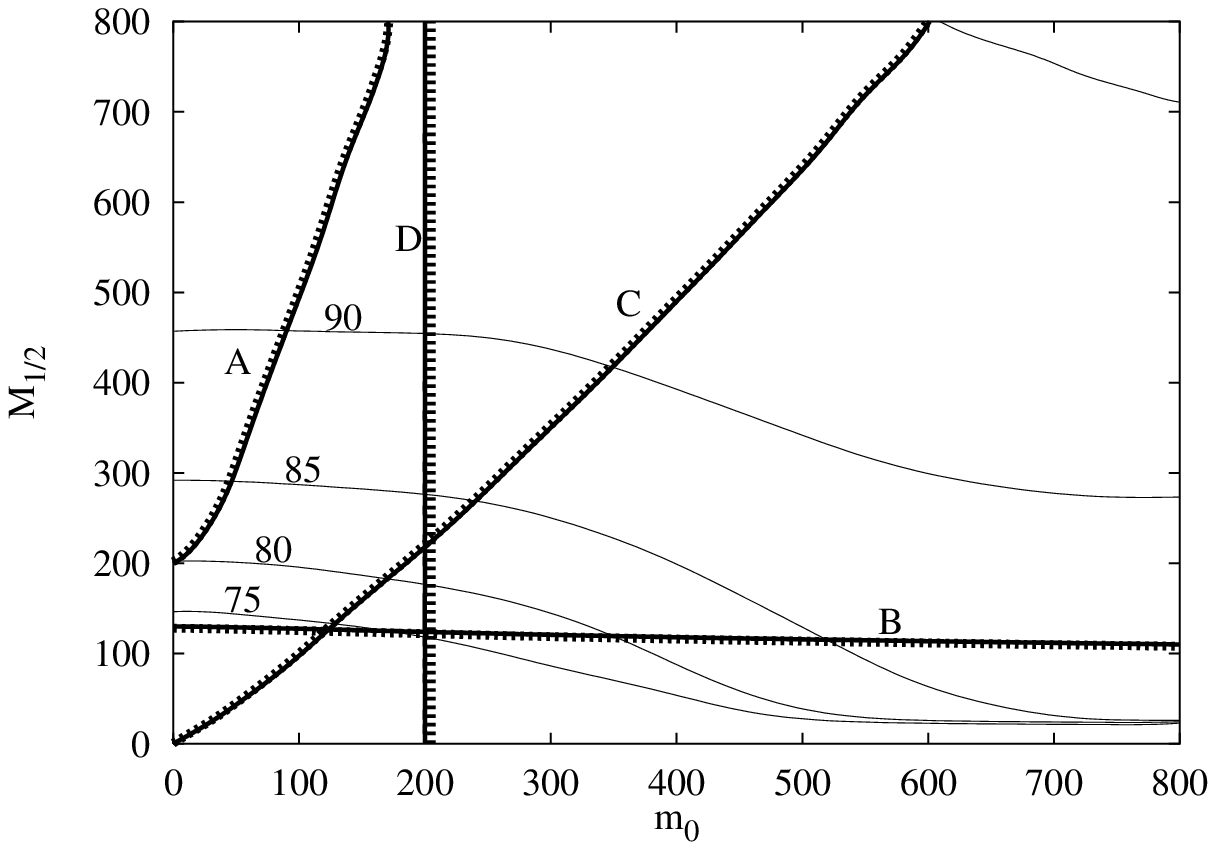}}
  \end{center}
  \caption{As in Fig.\ref{fail} but for $\mu <0$.}
  \label{failm} 
\end{figure}

We now apply some additional constraints to the ($m_0$,
$M_{1/2}$) parameter space.
Figs.~\ref{fail} and~\ref{failm} shows experimental bounds (see \eg
Ref.\cite{LEP2higgsbound}) and bounds from deep minima appearing in 
`unbounded from below' (UFB) directions 
in the potential~\cite{ccb,komatsu,casas,as}. Regions of parameter 
space above the line $M_{1/2}\geqsim m_0 $ have a minimum 
which can compete with the physical one and which is generally much larger. 
(The spectra we presented above were in regions of parameter space which 
violate this bound for reasons which will become apparent in the 
discussion.) 

The constraint is $m_0\geqsim 0.92 M_{1/2}$ at low $m_0$ and 
falls to $m_0\geqsim 0.75 M_{1/2} $ for larger values, mainly because 
of the larger values of $\mu $. 
This is in accord with the numerical 
work of Ref.\cite{casas}. The analytic (one loop) estimates of 
Ref.\cite{as} give $m_0\geqsim 1.12 M_{1/2}$ and 
$m_0\geqsim 0.95 M_{1/2}$ respectively and therefore represent an
overestimate of roughly 15-25\% in the bound.
The UFB bound was not included in the analysis of Ref.\cite{efos} 
and close to the QFP this is the severest bound.

The constraint that neutralino dark matter does not over close 
the universe should also be applied. The LSP should be 
able to annihilate quickly enough, for which we require that the 
masses of sparticles appearing in $s$ and $t$-channel processes be 
sufficiently small~\cite{dn}. This places a limit on the 
supersymmetry breaking scale; a full calculation is outside 
the scope of this paper and here we shall simply adopt the
overall limit found at the QFP in Ref.~\cite{bargerkao};
$m_0<200$ GeV. This is actually quite conservative; as may be seen 
for example in Refs.\cite{bargerkao,efos}; the tendency is 
for the dark matter bound to confine $M_{1/2}$ as well. 

We also impose that the neutralino is
the LSP (\ie lighter than the stau~\cite{subir}) and the 
chargino bound from LEP 2\cite{LEP2higgsbound}. (There are 
additional bounds coming from slepton searches at low $m_0$ 
which were not included here.)
The most restrictive experimental bounds are those from the
LEP2 lower bound on the standard model higgs mass. 
The CP-odd higgs $A^0$ is always much
heavier than the lightest CP-even higgs $h^0$, which results in the Standard
Model bounds being applicable to the quasi-fixed MSSM to good 
accuracy~\cite{espin}.
In Figs.~\ref{fail} and~\ref{failm} we show the light higgs contours for 
$m_{h^0} = 70, 75, 80, 85, 90,$ and $90\gev$. 
The latest lower bound from LEP 2 is $87\gev $~\cite{LEP2higgsbound}, 
but this is expected to rise. Even this
bound rules out both $\mu >0$ and $\mu <0$ when combined with the 
above constraints, {\em unless}\/ we allow the physical vacuum 
to be meta-stable~\cite{as} and/or ignore the dark matter bound, 
perhaps because of thermal inflation~\cite{thermalI}. 

\section*{Discussion}

We have presented the spectrum for the constrained MSSM
at the low $\tan \beta $ fixed point and have found that the model can 
only satisfy higgs and dark matter bounds in regions 
of parameter space where the physical vacuum is meta-stable. 
We should interpret this fact carefully since it does not necessarily exclude the model.
To see why, let us first clarify what the UFB bounds mean by 
summarising the conclusions of Refs.\cite{as,riotto}. 

The dangerous charge and colour breaking minima which lead to the UFB bounds 
form radiatively along $F$ and $D$-flat 
directions. However the vacuum decay rate is suppressed by a large 
temperature dependent barrier and the quantum tunneling rate 
is insignificant except at very small values of $m_0$. Thus a meta-stable vacuum 
would have survived until the present day. In addition the decay rate 
out of a meta-stable charge/colour breaking minimum  
back to the physical vacuum is also very small. Since vacuum decay 
is ruled out in either direction, the question of meta-stability is probably 
only of psychological relevance although, rather mystifyingly, it remains
the commonly accepted criterion. In Ref.\cite{as} it was suggested that 
a sufficient condition, that the only minimum be the physical one, 
is the bound we should use rather than the `traditional' UFB bound. 
However it was also shown that this condition is numerically 
very close to the `traditional' UFB bound so that all that is required is 
a change of emphasis; the correct interpretation is that regions of 
parameter space which violate a UFB bound have dangerous minima 
(global {\em or local}\/) which can compete with the physical one whereas those which 
satisfy the bound don't. If the UFB bound is violated, 
one is obliged to explain how the universe ends 
up in the physical vacuum and not in the charge/colour breaking 
one which is generally more `likely' (in that it is at least 
$10^3$ times wider than the physical vacuum). Some cosmological 
suggestions have been made in Ref.\cite{riotto} and references therein
although none have been worked through in great detail. (They 
may also entail making assumptions about cosmology, such as a high re-heat 
temperature, which may be at odds with nucleosynthesis for example~\cite{as}.)

We favour an alternative remedy for this model which is simply to add 
a small amount of $R$-parity violation. This would be
enough to make the LSP unstable, while still evading current experimental
bounds upon the magnitude of $R$-violation~\cite{herbster}. 
In this case the dark-matter 
bounds and the sneutrino-as-LSP bound vanish, although obviously we have to look 
elsewhere for a dark matter candidate. In addition the UFB bounds disappear 
for the reasons discussed in Ref.\cite{as}. Specifically, there are 
five dangerous UFB directions which correspond to the sets of invariants~\cite{komatsu,casas}, 
\ba
L_iH_2 &&  L_i L_3 E_3 \mbox{\hspace{2cm}; $i=1,2$}  \nn\\
L_iH_2 &&  L_i Q_3 D_3 \mbox{\hspace{2cm}; $i=1,2,3$},
\ea 
which are absent from the $R$-parity invariant superpotential. 
To lift the flat direction we can add the following 
lepton number violating contribution to the superpotential, 
\be
W_{B}=\lambda_{ijk}L_i L_j E_k + \lambda'_{ijk} L_i Q_j D_k.
\ee 
These operators are enough to lift the flat directions 
provided that they satisfy~\cite{as};
\ba
\label{lam3}
\lambda \geqsim \frac{0.26 h_\tau M_{1/2}}{\mu}
\approx 0.007 \frac{M_{1/2}}{\mu}\nn\\
\lambda' \geqsim \frac{0.26 h_b M_{1/2} }{\mu}\approx 
0.009 \frac{M_{1/2}}{\mu}.
\ea
(Here we find that $\mu/M_{1/2} \geqsim 1$.)
A suitable selection of non-zero $R$-parity violating couplings 
is $\lambda_{123}, \lambda'_{113},\lambda'_{223}, \lambda'_{333}\neq 0$
although other combinations are possible.
In Ref.~\cite{as}, it was shown that, provided they satisfy Eq.\ref{lam3},
these four couplings lift all five 
would-be UFB directions so that there are no {\em local}
\/minima except the physical one, whilst simultaneously evading 
the experimental limits~\cite{herbster} on $\lambda_{ijk},  \lambda_{ijk}'$.
They are small enough however that they will not significantly effect 
the spectrum in Figs.\ref{fail},\ref{failm}.

The CMSSM near the QFP is an attractive model in which the sparticle spectrum
depends upon only two parameters (modulo a choice of the sign of $\mu$). 
Models such as SUSY GUTs (that have the
MSSM as the effective field theory below $M_{GUT}$) which predict
bottom-tau Yukawa unification~\cite{btau} favour the QFP, and we
have shown this with a simple analytic argument.
For $m_t=175$ GeV, the model must be near the QFP~\cite{waoi}.

However the CMSSM at the low $\tan\beta $ fixed point
is ruled out by either recent higgs mass
bounds or dark matter constraints or the presence of a global UFB minimum.
Possible solutions include just living with
meta-stable vacuum or adding a small amount of 
$R$-parity violation~\cite{as}. 
$R$-parity violation can be made small enough to evade experimental
bounds in extensions of the MSSM~\cite{rpvnatural} without invoking 
very small fundamental dimensionless couplings. In addition the mass 
spectrum we presented would not change appreciably and hence the 
$R$-parity violating fixed point scenario
is still ultimately falsifiable due to the absolute upper bound upon the
higgs mass~\cite{espin} $m_{h^0} < 97\pm2$ GeV.
The measurement of the mass of one identified SUSY particle ought to be enough
to determine the entire sparticle spectrum (to within a discrete choice 
of the sign of $\mu$) near the QFP\@.

\newpage

\section*{Appendix}
Here, we present the analytic solutions 
to the one loop RGEs (see for example \cite{martin}) for the soft terms
of the MSSM, with arbitrary boundary conditions, in terms of
the three parameters with QFPs, $A_t$, $R$ and $M^2$.
They may easily be found without having to solve explicitly 
for $A_t$, $R$ and $M^2$. (See \eg Ref.\cite{as} and references therein 
for these solutions). Defining
\ba
\delta_i^{(n)} & = & 
(\alpha^n_i-\alpha_i^n|_0))/\alpha_i^n|_0 \nonumber \\ 
G &=& \frac{R}{R_0}
\left( \frac{\alpha_{3}}{\alpha_3|_0}\right)^{-7/9}
\left( \frac{\alpha_{2}}{\alpha_2|_0}\right)^{3}
\left( \frac{\alpha_{1}}{\alpha_1|_0}\right)^{13/99},
\ea
where the $0$-subscript indicates values at the GUT scale, the solutions
are
\ba
A_{U_{ij}}-\smallfrac{1}{2}A_t &=&
M_{1/2}\left(-\smallfrac{8}{9}\delta_3^{(1)}+
\smallfrac{3}{2} \delta_2^{(1)} +\smallfrac{13}{99} \delta_1^{(1)}\right)
+ 
(A_{U_{ij}}-\smallfrac{1}{2}A_t)|_0\nonumber \\
A_{U_{i3}}-A_t &=& (A_{U_{i3}}-A_t)|_0 G^{1/6}\nonumber \\
A_{U_{3j}}-A_t &=& (A_{U_{3j}}-A_t)|_0 G^{1/12}\nonumber \\
A_{D_{\alpha 3}}-\smallfrac{1}{6}A_t &=&
M_{1/2}\left(-\smallfrac{40}{27}\delta_3^{(1)}+
\smallfrac{5}{2} \delta_2^{(1)} +\smallfrac{29}{99} \delta_1^{(1)}\right)
+ 
(A_{D_{\alpha 3}}-\smallfrac{1}{6}A_t)|_0\nonumber \\
A_{D_{\alpha j}} &=& M_{1/2}\left(-\smallfrac{16}{9}\delta_3^{(1)}+
3 \delta_2^{(1)} +\smallfrac{7}{99} \delta_1^{(1)}\right) + 
(A_{D_{\alpha j}})|_0\nonumber \\
A_{E_{\alpha \beta}} &=& M_{1/2}\left(
3 \delta_2^{(1)} +\smallfrac{3}{11} \delta_1^{(1)}\right) + 
(A_{E_{\alpha \beta}})|_0\nonumber \\
B-\smallfrac{1}{2}A_t &=& M_{1/2}\left(\smallfrac{16}{9}\delta_3^{(1)}+
\smallfrac{3}{2} \delta_2^{(1)} +\smallfrac{5}{66} \delta_1^{(1)}\right) + 
(B-\smallfrac{1}{2}A_t)|_0\nonumber\\
m_{U_{33}}^2-M^2 &=& M^2_{1/2}\left(\smallfrac{8}{27}\delta_3^{(2)}+
\delta_2^{(2)} -\smallfrac{1}{27} \delta_1^{(2)}\right) + 
(m^2_{U_{33}}-M^2)|_0\nonumber\\
m_{Q_{33}}^2-\smallfrac{1}{2}M^2 &=&
M^2_{1/2}\left(\smallfrac{16}{27}\delta_3^{(2)}-
\delta_2^{(2)} +\smallfrac{5}{297} \delta_1^{(2)}\right) + 
(m^2_{Q_{33}}-\smallfrac{1}{2} M^2)|_0\nonumber\\
m_2^2-\smallfrac{3}{2}M^2 &=&
M^2_{1/2}\left(-\smallfrac{8}{9}\delta_3^{(2)}+
\smallfrac{2}{99} \delta_1^{(2)}\right) + 
(m^2_2-\smallfrac{3}{2}M^2)|_0\nonumber\\
m_{U_{3j}}^2 &=& (m^2_{U_{3j}})|_0 G^{1/6}\nonumber\\
m_{U_{i3}}^2 &=& (m^2_{U_{i3}})|_0 G^{1/6}\nonumber\\
m_{Q_{3j}}^2 &=& (m^2_{Q_{3j}})|_0 G^{1/12}\nonumber\\
m_{Q_{i3}}^2 &=& (m^2_{Q_{i3}})|_0 G^{1/12}\nonumber\\
m_{L_{\alpha\alpha}}^2 &=& M^2_{1/2}\left(-\smallfrac{3}{2}\delta_2^{(2)}-
\smallfrac{1}{22} \delta_1^{(2)}\right) +
(m^2_{L_{\alpha\alpha}})|_0\nonumber\\
m_{1}^2 &=& M^2_{1/2}\left(-\smallfrac{3}{2}\delta_2^{(2)}-
\smallfrac{1}{22} \delta_1^{(2)}\right) +
(m^2_{1})|_0\nonumber\\
m_{U_{ii}}^2 &=&  
M^2_{1/2}\left(\smallfrac{8}{9}\delta_3^{(2)}-
\smallfrac{8}{99} \delta_1^{(2)}\right) +(m^2_{U_{ii}})|_0\nonumber\\
m_{Q_{ii}}^2 &=& 
M^2_{1/2}\left(\smallfrac{8}{9}\delta_3^{(2)}-\smallfrac{3}{2}\delta_2^{(2)}
-\smallfrac{1}{198} \delta_1^{(2)}\right) +(m^2_{Q_{ii}})|_0 \nonumber\\
m_{D_{\alpha\alpha}}^2 &=& 
M^2_{1/2}\left(\smallfrac{8}{9}\delta_3^{(2)}-
\smallfrac{2}{99} \delta_1^{(2)}\right) +(m^2_{D_{\alpha\alpha}})|_0
\nonumber\\
\mu &=& \mu |_0 G^{1/4} 
\left(\frac{\alpha_3}{\alpha_3|_0}\right)^{-1}
\left(\frac{\alpha_2}{\alpha_2|_0}\right)^{-3/2}
\left(\frac{\alpha_1}{\alpha_1|_0}\right)^{-1/22}
\ea
where $ij = 1,2$ and $\alpha = 1,2,3$, and where we assume universal 
gaugino mass ($M_{1/2}$) at the high scale. (The solutions for the 
off-diagonal terms 
are only valid in a generic basis, \eg not the mass basis.) 
The remaining terms do not run in this approximation. A more 
general set of solutions (valid in any basis) for the flavour 
changing terms was presented in Ref.\cite{brax}.
 
\section*{Acknowledgements}
We would like to thank J.~Ellis, J.~Espinosa, T.~Falk, 
L.~Roszkowski, S.~Sarkar and C.~A.~Savoy for discussions.
BCA would like to thank CERN for hospitality extended during which part of
this work was carried out.

\end{document}